\documentclass[preprint,superscriptaddress,nofootinbib]{revtex4-1}
\usepackage{subfigure}
\usepackage{amsmath,graphicx,amssymb,float}
\usepackage{multirow,color}
\usepackage{url}
\usepackage{natbib}

\usepackage{bm,verbatim}

\renewcommand{\vec}[1]{\bm{#1}}

\newcommand{\eps}{\varepsilon}

\newcommand{\Rl}{R_{\lambda}}

\newcommand{\beq}{\begin{equation}}
\newcommand{\eeq}{\end{equation}}

\newcommand{\moritz}[1]{\textcolor{black}{#1}}

\begin{document}

\title{Self-organization and transition to turbulence in isotropic fluid
motion driven by negative damping at low wavenumbers}

\author{W.~D. McComb}
 \affiliation{
 SUPA, School of Physics and Astronomy,
 University of Edinburgh, UK}
 \author{M.~F. Linkmann}
 \affiliation{
 SUPA, School of Physics and Astronomy,
 University of Edinburgh, UK}
 \author{A. Berera}
 \affiliation{
 SUPA, School of Physics and Astronomy,
 University of Edinburgh, UK}
\author{S.~R. Yoffe}
 \affiliation{
 SUPA, Department of Physics,
 University of Strathclyde, Glasgow, UK}
 \author{B. Jankauskas}
 \affiliation{
 SUPA, School of Physics and Astronomy,
 University of Edinburgh, UK}

\begin{abstract}

We observe a symmetry-breaking transition from a turbulent  to a
self-organized state in direct numerical simulation of the Navier-Stokes
equation at very low Reynolds number. In this self-organised state the
kinetic energy is contained only in modes at the lowest resolved 
wavenumber, the skewness vanishes, and visualization of the flows shows 
a lack of small-scale structure, with the vorticity and velocity vectors
becoming aligned (a Beltrami flow).\\

\end{abstract}


\maketitle
One of the better known results in the history of science is the
discovery of the laminar-turbulence transition in pipe flow by Osborne
Reynolds in the late nineteenth century. Since then such transitions
have been found in many other flow configurations, and the subject is
widely studied today. In contrast, the commonly accepted view of
randomly forced isotropic fluid motion is that the motion is turbulent
for all Reynolds numbers and no actual transition to turbulence occurs.

In the course of studying the direct numerical simulation (DNS) of
forced isotropic turbulence, we have found that self-organized states
form at low Taylor-Reynolds numbers, in simulations which were invariably
turbulent at large Taylor-Reynolds numbers, thus hinting at the possibility
of a transition to turbulence. We observed depression of the
nonlinear term in the Navier-Stokes equation (NSE) during the formation
of the self-organized state. Visualization of the flow showed that in 
the self-organized state the velocity field $\vec{u}$ and vorticity field 
$\vec{\omega}= \nabla \times \vec{u}$
are aligned in real space, forcing the nonlinear term $\vec{u} \times
\vec{\omega}$ in the NSE to vanish. This is known as a Beltrami field 
\cite{Constantin88,Kambe10}. As it is also a condition for maximum helicity,
 and the initial state has zero helicity, it follows that the
transition to the self-ordered state is symmetry-breaking. Moreover, the
flow shows only large-scale structure. 

We begin by discussing the details of our `numerical experiment'.
The incompressible forced NSE was solved
numerically, using the standard fully de-aliased pseudospectral method 
on a 3D periodic domain of length $L_{box}=2\pi$, resulting in the lowest 
resolved wavenumber $k_{min}=2\pi/L_{box}=1$.
We note that our simulation method is both conventional and widely used, so
we give only some necessary details here. Full details of our numerical 
technique and code validation may be found in \cite{Yoffe12}.

We found that a choice of $32^3$ collocation points proved to be
sufficient  in order to resolve the Kolmogorov dissipation scale, as all
our simulations in this particular investigation were at very low Reynolds 
numbers. We did however verify that the formation of the self-organized 
state was also observed at a higher resolution ($64^3$).
All simulations were well resolved, satisfying $k_{max}\eta \geqslant 2.16$ 
\cite{McComb01a}, where $\eta$ denotes the Kolmogorov dissipation scale. 
The maximum time the simulations were evolved for was $t=1000s$, or, in terms 
of initial large-eddy turnover times $t_0=L/U$, $t=1271t_0$, where $U$ denotes 
the initial rms velocity and $L$ the initial integral length scale. 
Initial Taylor-Reynolds numbers $\Rl$ ranged from $\Rl=2.61$ to $\Rl=3.78$, where
$\lambda$ denotes the Taylor microscale.  
A summary of simulation details is given in Table \ref{tbl:simulations}.

\begin{table}[!t]
 \centering
 \begin{tabular}{ccccccc}
   $N^3$  & $k_{max}\eta$ & $R_{\lambda,SO} $  &  $\nu $ & $R_{\lambda}$ & $t_{max}/s$ \\
  \hline
  \hline
   $32^3$ & 2.85 & 12.91 & 0.1 & 2.61 & 1000   \\
   $32^3$ & 2.63 & 15.12 & 0.09 & 2.93 & 1000   \\
   $32^3$ & 2.56 & 15.91 & 0.087 & 3.00 & 1000   \\
   $32^3$ & 2.52 & 16.47 & 0.085 & 3.07 & 1000  \\
   $32^3$ & 2.48 & 17.07 & 0.083 & 3.14 & 1000   \\
   $32^3$ & 2.41 & 18.04 & 0.08 & 3.26 & 1000   \\
   $32^3$ & 2.39 & 18.38 & 0.079 & 3.30 & 1000   \\
   $32^3$ & 2.34 & 19.11 & 0.077 & 3.39 & 1000  \\
   $32^3$ & 2.29 & 19.88 & 0.075 & 3.47 & 1000   \\
   $32^3$ & 2.25 & 20.70 & 0.073 & 3.57 & 1000   \\
   $32^3$ & 2.29 & 21.58 & 0.071 & 3.67 & 1000  \\
   $32^3$ & 2.18 & 22.04 & 0.07 & 3.72 & 1000  \\
   $32^3$ & 2.16 & 22.52 & 0.069 & 3.78 & 1000   \\
   $64^3$ & 5.26 & 15.12 & 0.09 & 2.93 & 1000 \\ 
   $64^3$ & 4.82 & 18.04 & 0.08 & 3.26 & 1000 \\ 
   $64^3$ & 4.36 & 22.04 & 0.07 & 3.72 & 1000 \\ 
 \hline
  \end{tabular}
 \caption{Specifications of simulations. $N^3$ denotes the number of collocation points,
 $k_{max}$ the largest resolved wavenumber, $\eta$ the Kolmogorov microscale,
 $R_{\lambda,SO}$ the Taylor-Reynolds
 number in the self-organized state, $\nu$ the kinematic viscosity, $R_{\lambda}$
 the initial Taylor-Reynolds number and $t_{max}$ the time the
 simulations were evolved for.
}
 \label{tbl:simulations}
 \end{table}

The initial conditions were random (Gaussian) velocity fields with 
prescribed energy spectra of the form 
\begin{equation}
  E(k)=C_1k^4 e^{-2(k/k_0)^2} \ ,
\end{equation} 
where $C_1= 0.001702$ and $k_0=5$. The initial helicity 
$\langle \vec{u}(\vec{k})\vec{\omega}(-\vec{k}) \rangle$, where 
$\vec{\omega}(\vec{k})$ is the Fourier transform of the vorticity field 
$\vec{\omega}(\vec{x})$, was negligible for all simulations.

The system was forced by negative damping, with the Fourier transform of the 
force $\vec{f}$ given by
\begin{align}
 \vec{f}(\vec{k},t) &=
      (\varepsilon_W/2 E_f) \vec{u}(\vec{k},t) \quad
\text{for} \quad  0 < \lvert\vec{k}\rvert < k_f ; \nonumber \\
  &= 0   \quad \textrm{otherwise},
\label{forcing}
\end{align}
where $\vec{u}(\vec{k},t)$ is the instantaneous velocity field (in
wavenumber space). The highest forced wavenumber, $k_f$, was chosen to
be $k_f = 2.5$. As $E_f$ was the total energy contained in the forcing
band, this ensured that the energy injection rate was 
$\varepsilon_W =\textrm{constant}$.

Negative damping was introduced to turbulence theory by Herring
\cite{Herring65} in 1965 and to DNS by Machiels \cite{Machiels97a} in 1997. Since then
it has been used  in many different investigations to simulate
homogeneous isotropic turbulence (for example,
\cite{Jimenez93,McComb01a,Yamazaki02,Kaneda03,McComb03,McComb14b,McComb14c})
and particularly in the benchmark simulations of Kaneda and co-workers
\cite{Kaneda06}, with Taylor-Reynolds numbers up to $\Rl = 1200$. It has
also been studied theoretically by Doering and Petrov \cite{Doering05}. 

It should be emphasized that the forcing in the DNS does {\em not}
relate the force given in  \eqref{forcing} to the velocity field in the
NSE {\em at the same instant in time}.  The force added at time $t$ is 
calculated from the velocity field at the previous time step $t-dt$,
where in practice an intermediate predictor-corrector step is taken.
This has the important consequence that, because of nonlinear phase
mixing during this additional step, the force and the velocity field 
in the NSE at a given instant in time are not in phase, as long as the 
nonlinear term is active.  
Although this forcing procedure does not have an explicit stochastic
element,  it is nevertheless generally regarded as adding a random force
to the fluid. As the velocity field at $t=0$ consists of a
(pseudo-)randomly generated set of numbers, the forcing rescales  this
set of numbers and feeds it back into the system at the first time step;
and so on.  

\begin{figure}[tbp]
 \begin{center}
  \subfigure[ ~Evolution of the total energy]{
   \label{sfig:hit3d_E}
   \includegraphics[width=0.45\textwidth]{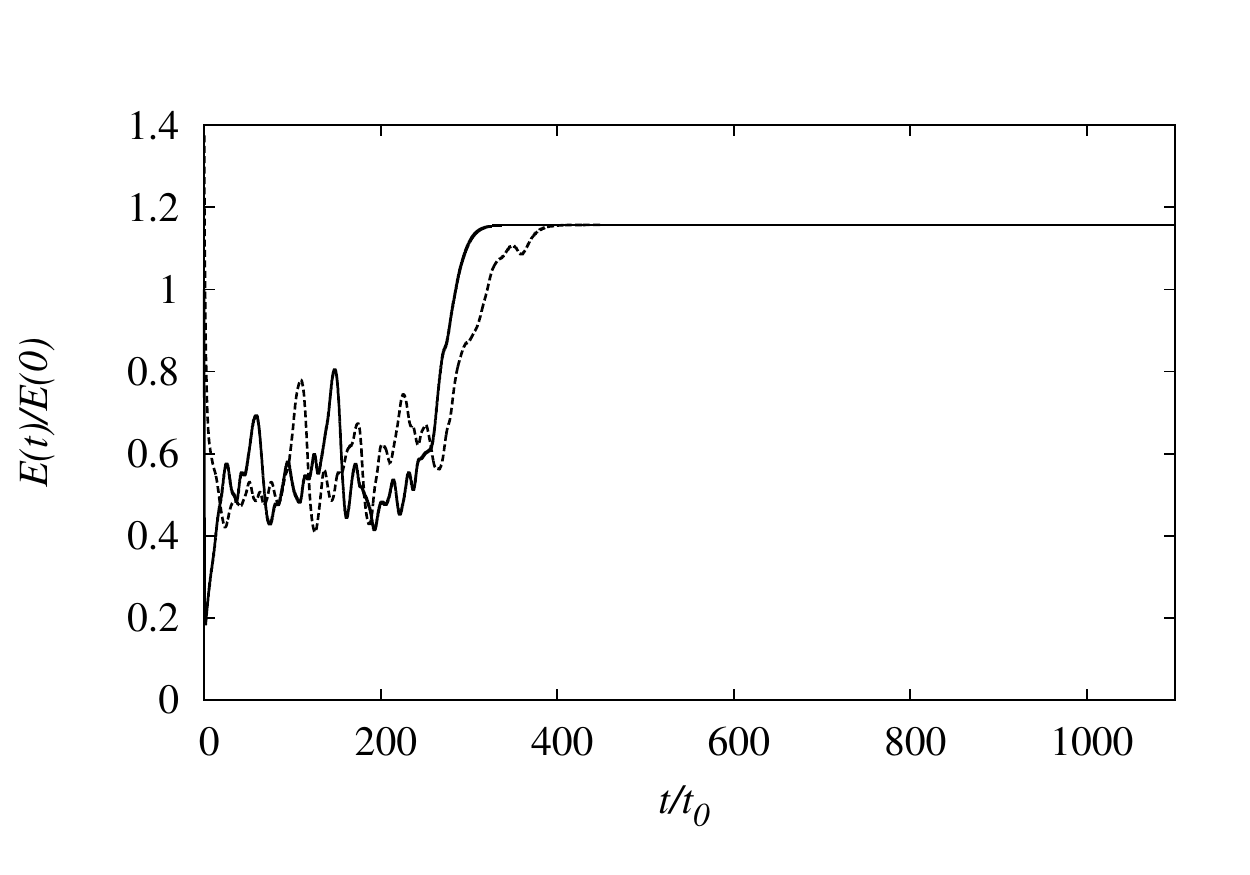}
   }
  \subfigure[ ~Evolution of the skewness]{
   \includegraphics[width=0.45\textwidth]{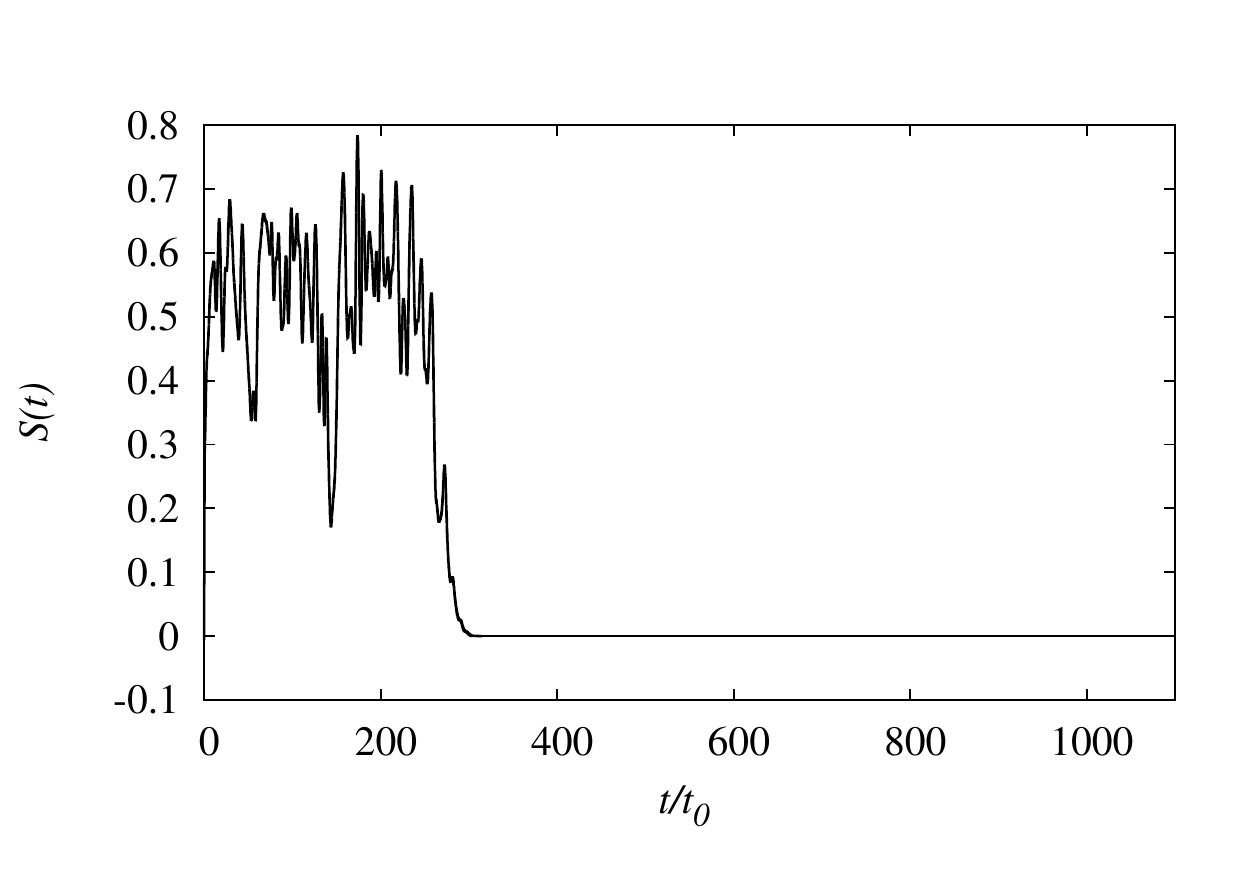}
   \label{sfig:ss_skew}
   }
 \end{center}
 \caption{Onset of the self-organized state as indicated by the total energy $E(t)$ and the 
          velocity-derivative skewness $S(t)$ for $\Rl= 3.39$.  
          (a) 
          $E(t)$: the solid line shows results from our DNS and the dashed line results 
          from {\tt{hit3d}}. \moritz{Note that the two realisations obtained from the different
          codes reach the same aysmptotic value of $E(t)$.} 
          (b) $S(t)$: evolution of the skewness for the same run. 
          Time is given in units of initial large eddy turnover time $t_0=L/U$, where $U$ is the 
          initial rms velocity and $L$ the initial integral scale. }

 \label{fig:hit3d}
\end{figure}

In order to investigate aspects of turbulence at high Reynolds number,
which  are beyond currently available computing power, our invariable
practice is to  keep $\eps_W$ constant, and reduce $\nu$. 
This corresponds to taking the limit of infinite Reynolds
number \cite{Batchelor53}.  We have carried out investigations up to
$\Rl=435$  with regards to the behaviour of the dimensionless
dissipation rate \cite{McComb14c} and the behavior of the second-order
structure function \cite{McComb14b}.  During these investigations, we
observed the peculiar asymptotic behavior of the fluid at low Reynolds
numbers as shown in Fig.~\ref{fig:hit3d}, which was confirmed using 
the publicly available code {\tt{hit3d}} \cite{Schumakov07, Schumakov08}. 

At such low Reynolds numbers, we found that initially the simulations
developed in the usual way, with clear evidence of nonlinear mixing. 
But, after long running times, the total energy ceased to fluctuate, and
instead remained constant, while the skewness dropped to zero. Once this
had happened,  the kinetic energy was confined to the $k=1$ mode 
\moritz{\footnote{\moritz{That is, for the six wavevectors 
$(\pm 1, 0,0),(0,\pm 1, 0)$ and $(0,0,\pm 1)$.}}} 
and the
 nonlinear transfer had become zero. The cascade process was thus absent
and no small-scale structures  were being formed. Hence the system had
self-organized into a large-scale state.

Similar results were found both with our code and with 
{\tt{hit3d}} as shown in Fig.~\ref{sfig:hit3d_E}. 
Figure \ref{sfig:ss_skew} shows the 
evolution of the velocity derivative skewness. Note
that it drops to zero at long times, indicating that the observed state is Gaussian 
and hence is not turbulent. As the velocity derivative skewness is not 
recorded by {\tt{hit3d}}, a comparison using this quantity was not possible. 

Since the energy spectrum in the self-organized  state is unimodal at
$k=1$, we can predict the asymptotic value $E(t)=E_{\infty}$ in this
state from the energy input rate  $\varepsilon_W$ and the viscosity $\nu$ 
\moritz{
using the spectral energy balance equation for forced isotropic turbulence 
\begin{equation}
 \frac{\partial E(k,t)}{\partial t} = T(k,t) - 2\nu_0 k^2E(k,t) + W(k,t) \ ,
\label{eq:Lin}
\end{equation}
where $E(k,t)$ and $T(k,t)$ are the energy
and transfer spectra, respectively, and
\beq
W(k,t) = 4\pi k^2 \langle \vec{u}(-\vec{k},t)\cdot\vec{f}(\vec{k},t) \rangle
\eeq
is the work spectrum of the stirring force. 
By invoking stationarity ($\eps_W=\eps$, where $\eps$ denotes the
dissipation rate), we} 
obtain for the total energy after
self-organization 
 \begin{equation}
   \label{eq:E_B}
   E(t) = E_{\infty}  =\frac{\varepsilon_W}{2\nu}=\mbox{ constant} \ .
 \end{equation}
All runs for all values of the initial Taylor-Reynolds number stabilized
at this value of $E(t)$. Note that the same value is obtained both with
our code and with {\tt{hit3d}} as shown in Fig.~\ref{sfig:hit3d_E}. 
\moritz{The realisations obtained from {\tt hit3d} and from our code only
 differ in the random initial condition while all other parameters such as the 
viscosity and forcing rate are identical. As can also be seen in Fig.~\ref{sfig:hit3d_E},
although both realisations attain the same asymptotic value of $E(t)$, they do 
not do so at the same rate.}

\begin{figure}[!h]
 \begin{center}
  \subfigure[~Transfer spectra at different times for $\Rl=3.39$.]{
   \includegraphics[width=0.45\textwidth]{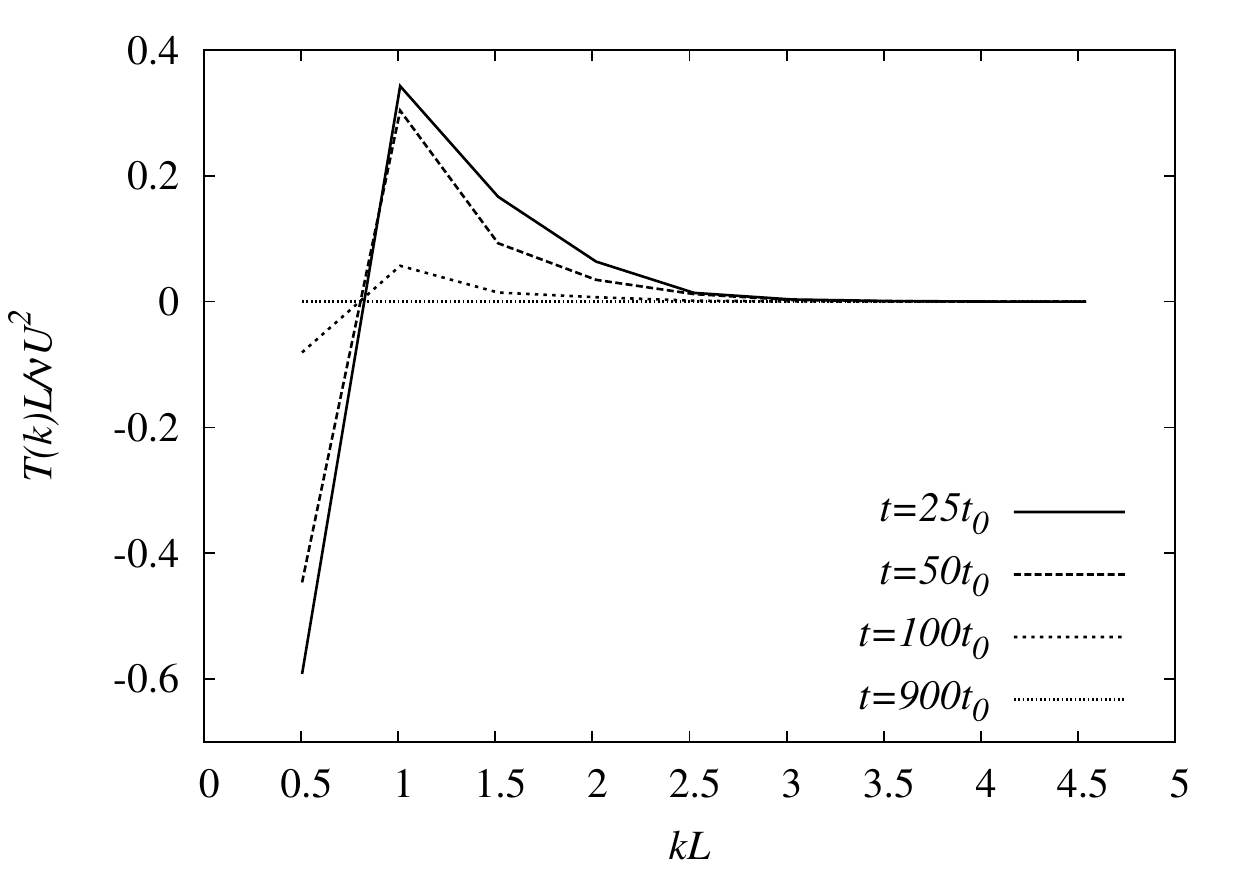}
    \label{sfig:transSO}
   }
  \subfigure[~Energy spectra at different times for $\Rl=3.39$.]{
   \includegraphics[width=0.45\textwidth]{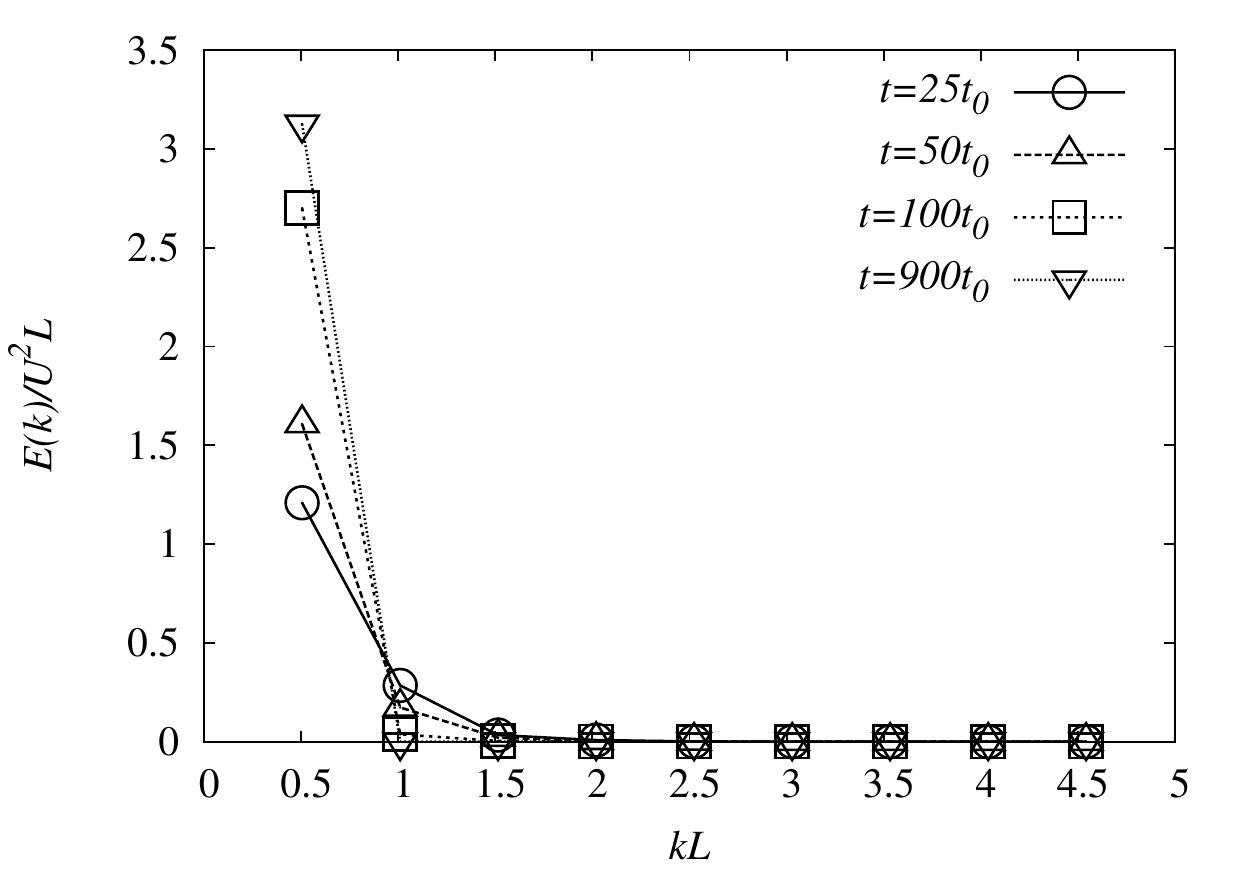}
    \label{sfig:specSO}
   }
 \end{center}
 \caption{
(a)  Transfer spectra at different times showing the depression of nonlinearity as the nonlinear transfer
    gets weaker as time progresses. The time evolution of the total energy corresponding to the measured
    spectra is shown in Fig.~\ref{sfig:hit3d_E}.
(b) Energy spectra at different times. The formation of a unimodal spectrum in clearly visible.
After self-organization only the $k=1$ mode remains populated while all other modes have
lost their kinetic energy, \moritz{where we note that $L\simeq 0.5$}. The kinetic energy is therefore concentrated in the large scales
and we do not expect small-scale flow patterns. This corresponds
to the right-hand image of Fig.~\ref{fig:visualisation}, where the flow shows only large-scale structure.
}
 \label{fig:evolution}
\end{figure}

The development of the large-scale state was accompanied by a depression
of nonlinearity, as shown in Fig.~\ref{sfig:transSO} where the transfer spectrum  
$T(k,t)$ is plotted at different times.  Again, it should be emphasized that {\em at early times
$T(k,t)$ has the behaviour typically found in simulations of isotropic
turbulence} (see \emph{e.g.}~Fig.~9.3, p.~293 of  \cite{McComb14a}), but
at later times tends to zero. At the same time, the energy spectrum
$E(k,t)$, tends to a unimodal spectrum at $k=1$ as can be seen in Fig.~\ref{sfig:specSO}. 

\begin{figure}[tbp]
 \begin{center}
   \includegraphics[width=\columnwidth]{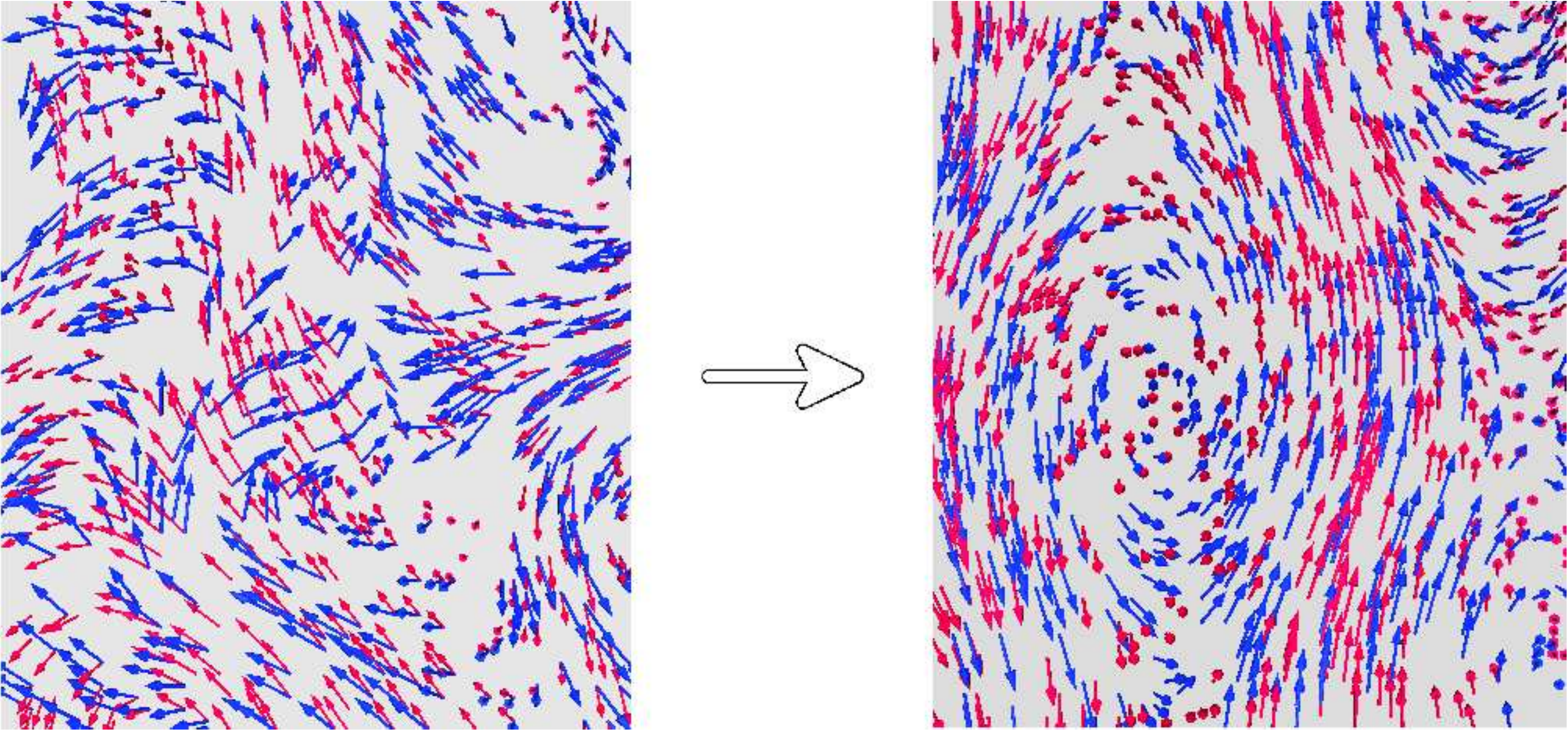}
 \end{center}
 \caption{(Color online) Visualization of the velocity field 
  (red arrows) and the vorticity field (blue arrows) in $x$-space 
  before and after self-organization. The alignment between velocity 
  and vorticity fields is clearly visible in the right-hand image. 
\moritz{The image shows a slice parallel to the $xy$-plane of the 
computational domain.}
}
 \label{fig:visualisation}
\end{figure}

Figure \ref{fig:visualisation}  shows snapshots of the system before and 
after self-organization. The arrows indicate the velocity field 
$\vec{u}(\vec{x})$ and vorticity field $\vec{\omega}(\vec{x})$ at  
different points in real space. We may see that the vorticity 
field and the velocity field align with each other once the system 
has self-organized, such that
\beq
\nabla \times \vec{u}(\vec{x})=\vec{\omega}(\vec{x})
                              = \alpha\vec{u}(\vec{x})\ ,
\label{eq:beltrami}
\eeq
for a coefficient $\alpha$ (which must have dimensions of inverse
length). Vector fields satisfying \eqref{eq:beltrami} are eigenfunctions
 of the curl operator and therefore helical. It should be noted that 
the nonlinear term in the NSE vanishes identically  if the velocity and
vorticity fields are aligned. 
From the image on the right hand side of Fig.~\ref{fig:visualisation}, it can
clearly be seen that the flow is in an ordered state, as velocity  and
vorticity vectors are everywhere parallel to each other, giving visual 
evidence of the Beltrami property \eqref{eq:beltrami} of the large-scale
 field. Furthermore, we observe a lack of small-scale structure of the
flow,  which reflects the measured unimodal energy spectrum in the
self-organized  state. The Beltrami condition has also been verified by
analyzing helicity spectra, showing nonzero helicity at $k=1$ and zero
helicity at all higher wavenumbers. \moritz{We found that the
final states always satisfied the Beltrami condition. 
However, final states occur where vorticity and velocity are aligned (positive 
helicity) or anti-aligned (negative helicity). 
Which of the two possible helicity states is chosen 
in the self-organized state depends on the random initial conditions.
If a particular final state occurs, then its mirror image will also 
be a possible asymptotic state that may occur, as there is no 
systematic preference for a particular configuration.    
} 

A short film showing the transition to the self-organized state can be
found in the supplemental material (available online at
stacks.iop.org/jpa/48/25FT01/mmedia). Two interesting points may 
be observed from the film. First, the system shows behaviour similar to 
`critical slowing down' (see e.g.~\cite{Hohenberg77}). Secondly, we 
observe a transition from a non-helical state to a
helical state. That is, the formation of the helical self-organized
state is symmetry-breaking. This latter aspect can also be seen in
Fig.~\ref{fig:visualisation}. 

Our results may be peculiar to our particular choice of forcing. We note that as
the nonlinear term tends to zero, the phase-mixing also vanishes, and the 
negative damping becomes in phase with the velocity field. Accordingly there is
then no mechanism to restore randomness, and in this simple picture, the 
self-organized state must be stable. \moritz{A further implication of this, 
is that negative damping would be rather
difficult to achieve in a real physical system. Accordingly we are
currently exploring the effect of other types of forcing and this work
will be reported in due course.}

However, it is arguable that the combination of the NSE and negative damping
is an interesting dynamical system in its own right. For initial values of the
Taylor-Reynolds number greater than about $25$ (corresponding to a steady-state value of 
about $40$), it reproduces isotropic turbulence; while for initial $\Rl$ less 
than about 5, it makes a spontaneous, symmetry-breaking transition to a 
Beltrami state. In the past, there have been 
mathematical conjectures about the possibility of Beltrami fields 
\cite{Moffatt85,Moffatt92} (and references therein) in turbulence; and 
numerical studies have suggested that such fields can occur in 
a localized way \cite{Kraichnan88,Pelz85,Kerr87}. 
To the best of our knowledge, this present investigation reports the first 
instance of a spontaneous transition to a stable Beltrami state for an entire flow field.

Furthermore, the emphasis in DNS has always been on 
achieving ever-larger Reynolds numbers, in order to study asymptotic scaling
behaviour. Yet in the case of the dimensionless dissipation, for example, 
the emphasis is on low Reynolds numbers, and indeed on looking for 
universal behaviour under these circumstances. Accordingly, the present 
work also draws attention to the surprising fact that stirred fluid motion
at low $\Rl$ may not actually be turbulent. 

One of the few practical applications of isotropic turbulence 
is in geophysical flows. Our results could possibly be relevant to atmospheric 
phenomena such as `blocking' \cite{Frederiksen07}, for instance.
Recent numerical studies in magnetohydrodynamics show the development of
similar organized states \cite{Dallas15}. This suggests a wider relevance of this type of result, not least
because it was achieved in a different system using a different forcing scheme from our own.

Investigation into the nature of this transition continues.
In view of recent successes of a dynamical systems approach to 
the transition to turbulence in wall-bounded shear flows, work is currently
being carried out taking a similar approach, and this will be submitted for 
publication in due course.

\section*{Acknowledgments}

This work has made use of computing resources provided by 
the Edinburgh Compute and Data Facility 
({\tt http://www.ecdf.ed.ac.uk}). A.~B. is supported by STFC, S.~R.~Y. and M.~F.~L. are funded by 
\moritz{the UK Engineering and Physical Sciences Research Council (EP/J018171/1 and EP/K503034/1)}.
Data is publicly available online \cite{sos_data}.

\bibliography{wdm_sos,pipe}	 

\end{document}